# Anapole resonances in low-index materials

Nicholas Joel Damaso, Sejeong Kim*

*School of Mathematical and Physical Sciences, University of Technology Sydney, New South Wales, 2007, Australia*

*\*Sejeong.Kim-1@uts.edu.au*

**Abstract**

Photonic cavities are valued in current research owing to the multitude of linear and nonlinear effects arising from densely confined light. Cavity designs consisting of low loss dielectric materials can achieve significant light confinement, competitive with other schools of cavity design such as plasmonics. However, the basic concepts in all dielectric photonics such as anapole resonances in nanodisks have been primarily studied in high index materials such as $WS_2$ and Si. Without additional measures, low index dielectric nanodisks struggle to achieve similar levels of light confinement. Here, we present fabricable design space for higher confinement in a low index dielectric cavity by incorporating the extensively studied, isolated dielectric nanodisk into broader host structures. In particular, we focus on hexagonal boron nitride (hBN), a novel dielectric 2D material with bright, room temperature single photon emitters and refractive indices of 2.1 and 1.8 in the in-plane and out-of-plane directions. Due to hBN's potential as a quantum light source, we characterise our cavities by their achievable Purcell factors at the anapole resonance. The effects of the supporting structures on the cavity resonances include boosts to the Purcell factor by as much as three-fold to a maximum observed factor of 6.2.

**Introduction**

The optical cavity is a particularly diverse tool in nanophotonics, enabling fundamental study of light-matter interactions in the weak and a strong coupling regimes as well as having the potential for practical implementations in an integrated photonic chip. Both dielectric and plasmonic cavities have been intensively studied in this regard for the past few decades [1, 2]. Dielectric cavities, which encompass photonic crystal cavities and micro disks/rings, are promising due to their high achievable Q-factors (Q) [3], however, they demand device sizes that are larger than a few micrometres to form a photonic band gap or to support total internal reflection of the whispering gallery mode (WGM). In contrast, plasmonics presents alternative approach where light can be confined in a nanoscale sized gap between two metal nanostructures. Consequently, the typically low-Q plasmonic resonances have significantly reduced mode volumes and hence, the achievable Purcell factors of these cavities can remain high due to the well-known proportionality with (Q/V). However, the utility of plasmonic cavities is currently limited by intrinsic metallic loss and non-CMOS compatible fabrication processes [4, 5].

The experimental observation of nanoscale resonances in silicon particles has spurred forward an active research effort into of these distinctly small footprint dielectric cavities [4-6]. Such cavities are called Mie resonators after the Mie theory which describes unique electromagnetic interaction between incident light and particles of a comparable size to the wavelength. The characteristic low loss environment of dielectrics has established a unique playground convening new and old electromagnetic effects, among them bound states in the continuum [7], directional scattering [8-10], magnetic resonances [11], second and third harmonic generation [12-14] and wavefront shaping metasurfaces [15, 16]. A particular avenue of research into the Mie resonator has opened since the discovery of the optical anapole in dielectric particles [17]. Anapole resonances in dielectric particles are especially non-scattering states, achieving lower total scattering efficiencies than even Rayleigh scattering particles [18]. Using the language of the Cartesian multipole expansion, the anapole phenomena is classified as an interference effect in the far field radiation associated with two irreducible charge/current distributions [19-22]. The light confinement of the anapole phenomenon has been deployed in high refractive index media such as silicon and GaAs for nonlinear and lasing applications [13, 23]. Recently, it was found that a TMDC material with high refractive index, namely $WS_2$, facilitated strong coupling in the formation of so-called anapole-exciton polaritons [24]. As of yet, the study of the Mie resonance's effect on the quantum emitters have not been investigated in detail.

Introducing low index particles to the list of useful anapole hosts is a current challenge [14]. The typical design of a quasi-anapole host cavity, the nanodisk, cannot excite the theoretical zero-scattering anapole but instead a leaky, quasi-anapole mode [25]. Attempting to use low index materials therefore increases the significance of this mode leakage and hence, hampers light confinement. Here, we study the anapole mode in a low refractive index material (n=2.1) and its effects on an internal dipole emitter via the Purcell effect. The refractive index of the cavity is that of hexagonal boron nitride (hBN) [26], a 2D material that contains single photon emitters at room temperature [27, 28]. To meet the challenge of improving light confinement beyond that of a typical nanodisk geometry, we investigated a potential suspended resonator and additionally, we tested different metal substrates and found that silver shows the highest Purcell which we attribute to the low real-part of its refractive index at visible wavelengths. These ground-level simulation results will serve as direction for future practical implementations of low index Mie resonances, especially for quantum photonics applications.

**Result and discussion**

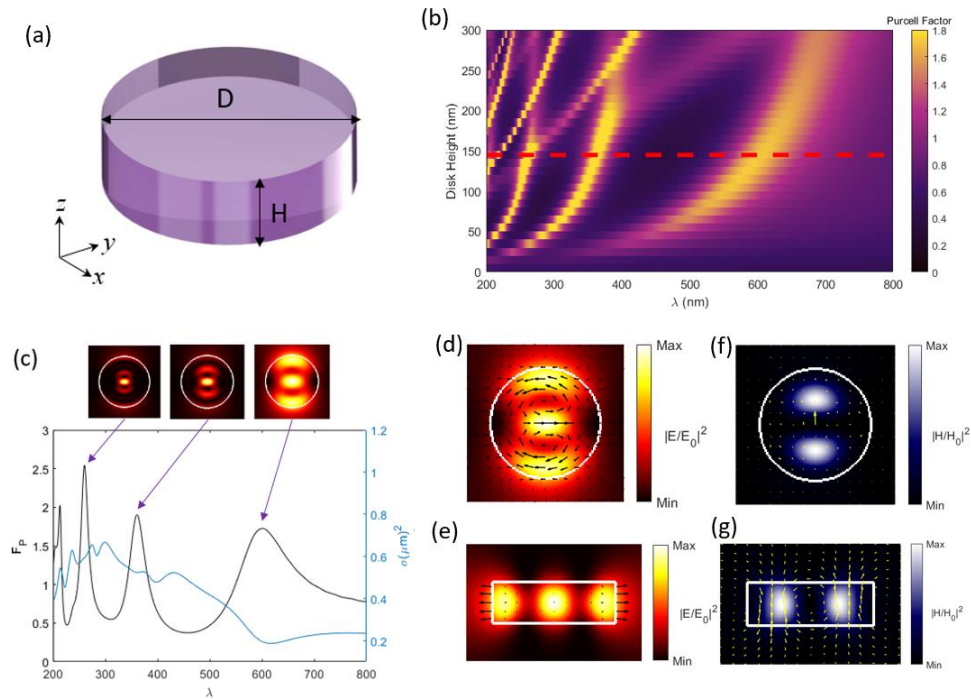

**Figure 1.** Free-standing hBN disk | (a) 3D Schematic of the dielectric disk. (b) 2D spectral map of the Purcell factor of a disk with varying height (H) with fixed diameter of 400 nm. Red dashed line (H=140 nm) indicates the dimension yielding the strongest resonance in the first-order Mie mode. (c) Purcell factor (Fp) and scattering cross section ($\sigma$) spectra obtained from dipole and plane-wave excitation respectively when the disk dimensions are D=424 nm and H=140 nm. Electric field intensity profiles of the first ($\lambda$ = 600 nm), second ($\lambda$ = 360 nm) and third ($\lambda$ = 260 nm) are shown. Local electric polarization map (black arrows) superimposed on cross-sections of the electric intensity in (d) x-y and (e) y-z plane are shown respectively. Local magnetic polarization map (yellow arrows) superimposed on the magnetic intensity as in (d,e).

We first start analysis of the isolated hBN nanodisk in a vacuum as shown in Figure 1a. An x-polarized dipole source is located at the centre of the disk, which is modelled with birefringent indexes of 2.1 for in-plane and 1.8 for the out-of-plane (z-axis) directions. Increasing the diameter (D) of the disk redshifts the resonant modes of the particle (Supplementary Information S1). Here we fix the diameter to 424 nm so that the dip to scattering in Figure 1c is located specifically at $\lambda$ = 600 nm, targeting the wavelength of the hBN single photon emitter. The disk height (H) sweep in Figure 1b shows that the mode at $\lambda$ = 600 nm is strongest when H is 140 nm.

A line scan of the Purcell map across the red-dashed line together with the scattering cross section is plotted in Figure 1c. The coincidence of a resonance in the Purcell factor and dip in the scattering cross section indicates that the enhancement stems from an anapole mode. Moreover, the internal electric and magnetic polarizations (Fig. 1d-1g) are in line with the anapole field profiles previously calculated and observed experimentally in nanodisks [21, 29]. Note that here the anapole wavelength is accompanied only by a minima instead of a zero in the scattering cross-section, this is indicative of the imperfect realization of the anapole as there is a presence of other multipole orders that remain radiative [21].

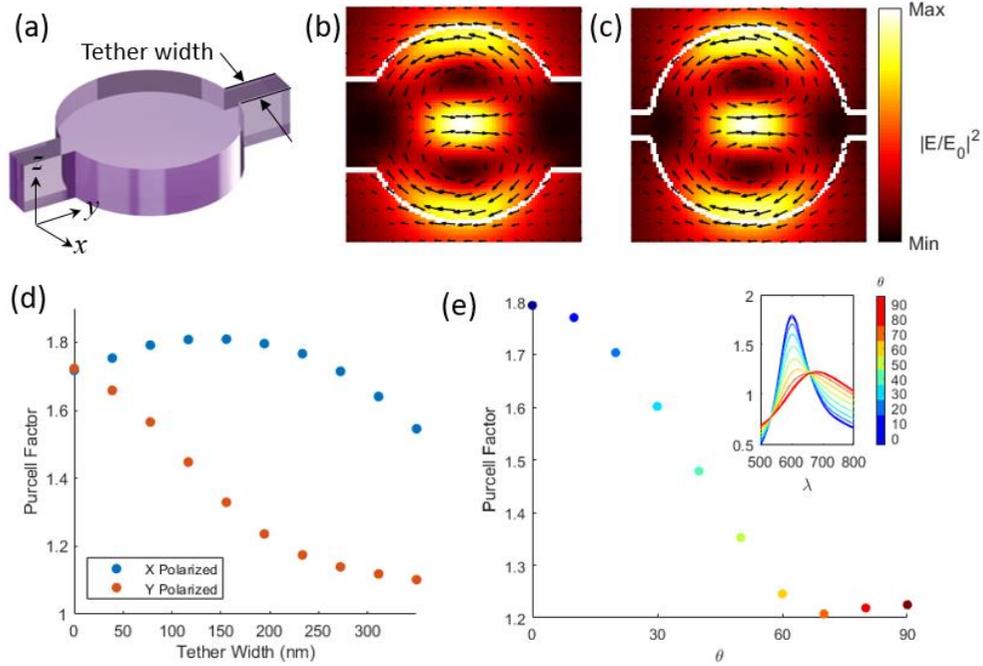

**Figure 2.** A microdisk with tethers | a) Schematic of the hBN disk with tether structure. b,c) Mode profiles of the electric field excited with a dipole source polarized parallel to the tether for (b) 200 nm and (c) 50 nm wide tethers. Black arrow indicate local electric polarization vectors. d) Purcell factors as a function of tether width when the electric dipole source is polarized x (blue dots) and y (red dots) e) Purcell Factors of the 200 nm wide tether cavity as a function of angle, $\theta$, of a dipole source, where $\theta = 0$ corresponds to the x axis. Inset shows Purcell spectra of each dipole angle $\theta$.

Initially, we propose the suspended structure (Fig. 2a) as a potential design for experimental fabrication. This morphology is similar to the kinds of devices previously developed in hBN photonic crystal cavities and may be produced by similar fabrication methods, namely a combination of a trenched substrate and electron beam lithography [30].

The anapole mode exhibits asymmetry dependent on the excitation polarization as seen in the perpendicular direction of the side lobes with respect to the dipole direction in Figure 2b. This asymmetry has been exploited in a previous study to selectively couple the near fields of the cavity to specific waveguides in the direction of the perpendicular lobes [13]. The same logic can be employed in the case of the tether. For the ideal emitter, with parallel polarization to the tether and centred position, the addition of the tether is seen to not significantly impact the mode at 600 nm (Fig. 2b, 2c) as the major lobes do not see the tether. Figure 2d shows the Purcell factors for the x- and y- polarized dipoles by varying the thickness of the tether. Here we focus on in plane emitters, as the resident emitters in the hBN crystal tend to be polarized in plane, however total control of emitter polarization is beyond current fabrication capabilities [31]. The Purcell factor of the y-polarized dipoles decrease quickly as the tether size increases while the x-polarized dipoles are barely affected by the tether. The Purcell spectra plotted for different in-plane polarization angles in Figure 2e demonstrate the extent to which the tether interferes with the mode, nearly totally nullifying the Purcell effect at non optimal angles.

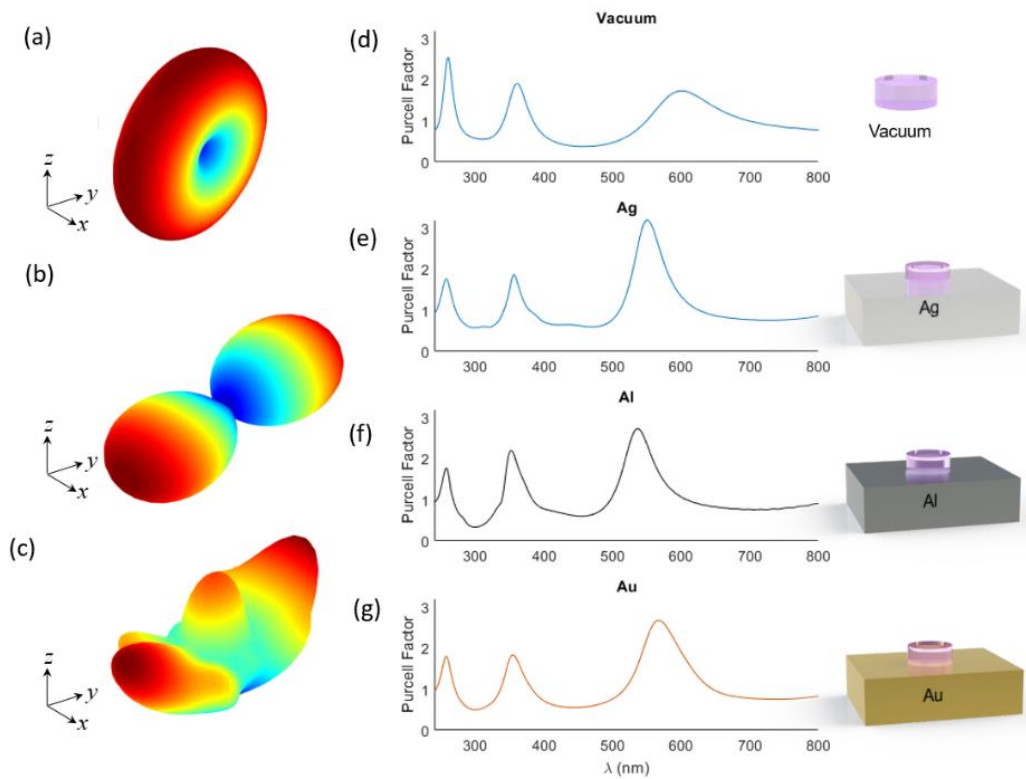

**Figure 3.** A Mie resonator on substrates | A 3D far field pattern comparison of a dipole emitter when the dipole is radiating in (a) free space, (b) in the free-floating nanodisk at the anapole resonance and (c) in the resonator on a silver substrate. d)

A reference Purcell spectrum from the free floating disk. e-g) Purcell spectra obtained from the disk placed on (e) silver, (f) aluminium and (g) gold substrates.

Despite Purcell enhancement of the tether structure relative to the free-floating disk, this enhancement is insufficient in making up for the low index material used and the Purcell factors remain too low for experimental observation. Another method of experimental realization of this Mie resonator is to place the disk on a substrate. However, introducing substrates with refractive indexes greater than that of a vacuum is conventionally known to degrade the Q-factor of the cavity due to decreased refractive index contrast at the substrate boundary. However, it was recently shown that low index materials such as aluminium or other metals with a refractive index less than 1 in the visible range can generate more efficient confinement in anapole resonances [14].

Here we analyse the effect of metal substrates on dipole emitters with respect to their far-field patterns and Purcell factors. Figure 3a shows the typical doughnut shaped far-field radiation pattern from a dipole emitter. The same dipole emitter shows enhanced directionality in its emission pattern when it is radiating through an anapole mode (Fig. 3b). This far-field is modified when the metal substrate is introduced, with most of the light emitted toward the upper hemisphere, which would facilitate increased collection efficiency. Figure 3d shows the Purcell factor with respect to the wavelength when the dipole emitter is located at the centre of the disk. From 3e to 3g, the modulation of the Purcell factor due to using different metal substrates is shown. Here, we note that the wavelength region of interests is around 600 nm considering the wavelength of hBN quantum emitters. The anapole, which had been previously tuned to 600, is enhanced by each of the substrates and spectrally blue shifted. The maximum enhancement is seen in silver, which achieves a Purcell factor of 3.2. We attribute this to silver's lowest part of the real refractive index at the spectral position of the shifted resonance (n = 0.28 at 550 nm) compared to aluminium (n = 0.83 at 536 nm) and gold (n = 0.29 at 567 nm).

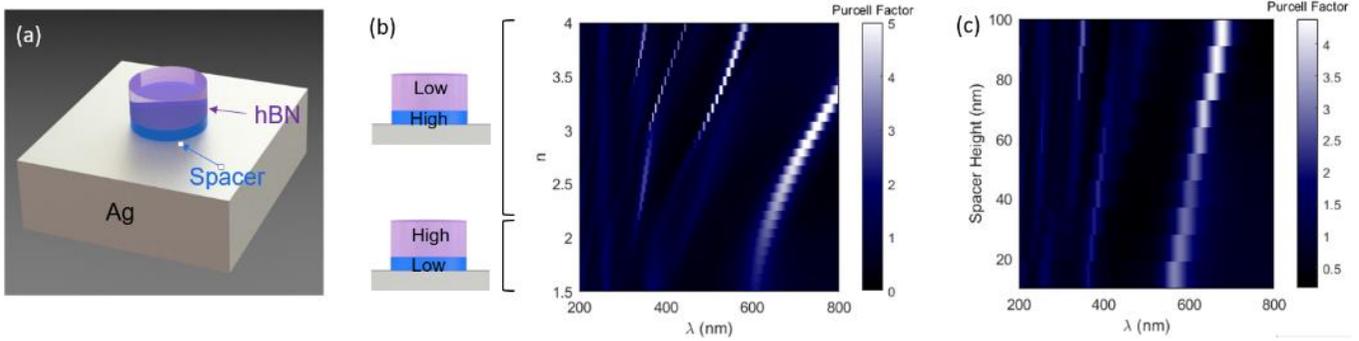

**Figure 4.** Adding dielectric spacer | a) Schematic of the Mie resonator with a spacer. The hBN disk defined previously sits atop a disk with equal radius and variable height and refractive index. The combined structure is seated on a silver substrate. b) 2D map of the Purcell factor of a dipole emitter at the centre of the hBN nanodisk in (a) atop a 100 nm thick spacer with varying refractive index. c) 2D Purcell Factor map using a spacer of n = 2.5 and varying height h.

The structure in Figure 4 is inspired by the practice within plasmonics of introducing a spacer to reduce unwanted interaction between emitters and a nearby metal. In our implementation, we model an arrangement of two stacked disks on a silver substrate and introduce a dipole emitter to the centre of the top disk which is made of hBN (Fig. 4a). The choice of silver is motivated by the superior Purcell factors achieved with the material in Figure 3. The structure serves to distance the hBN cavity mode from the ohmic losses in the substrate. Moreover, spacers with an index greater than that of hBN increase the contrast in refractive index between the cavity and substrate, further reducing transmission into the substrate. Indeed, the Purcell factor of the anapole mode (600 - 800 nm) increases monotonically with refractive index. In this case, the sacrifice of the monolithic nature of the cavity drives an enlargement of the achievable Purcell factor from 3.2 without the spacer (as in Figure 3) to 6.2 for a spacer with n = 3.3 (Fig 4b). In Figure 4c, an increasingly tall spacer enables ranging Purcell factors at the anapole resonance from 3.2 (substrate only) to 4.3 at 100 nm. That the refractive index of the spacer induces more significant enhancement than height suggests the relatively greater importance of index contrast compared to the actual dimensions of the spacer.

Increasing the refractive index and height of the spacer induce large red-shifts on the order of 200 nm to the mode at 600 nm in addition to other modes. This would justify a significant decrease in the cavity diameter to counteract the change, reducing the footprint of the cavity and potentially boosting the Purcell factor via a reduction in mode volume.

**Conclusion**

Movement towards utilizing anapole resonances to increase the brightness of hBN single photon emitters is limited by the material's relatively low refractive index. An hBN nanodisk hosting an anapole resonance, when inserted into a variety of supporting structures experiences increased Purcell factors at the anapole resonance by upwards of three-fold. The primary enhancement heuristic we have followed is the establishment of high contrasts in the refractive index across the boundaries of the nanodisk under the assumption that this reduces transmission through the boundary, hence reducing radiative losses from the cavity mode. Our all-hBN design therefore features little enhancement to the free space Purcell Factor while the substrate and high index spacer are seen to enhance the resonance, achieving Purcell Factors up to 6.2. An additional benefit of the substrate based designs is the rotational symmetry of these cavities, allowing equivalent modes to be excited from any excitation polarization.

Low index dielectrics other than hBN may be incorporated interchangeably in the central disk - pending tuning of the geometry. Lowering the bar for the applicability of dielectric materials eases design constraints and in addition, facilitates the enhancement of a wider range of photonic effects belonging to different materials.